\documentclass[nofootinbib,twocolumn,showpacs,amsmath,pra,aps,amssymb,superscriptaddress]{revtex4-1} 
\usepackage[T1]{fontenc}
\usepackage[latin9]{inputenc}
\usepackage{times}
\usepackage{color} 
\usepackage{xspace}
\usepackage{amssymb,amsmath}
\usepackage{amsbsy}
\usepackage[pdftex]{graphicx}
\usepackage{bm}
\usepackage{float}
\usepackage[normalem]{ulem}

\usepackage[unicode,breaklinks]{hyperref}
\hypersetup{
    unicode=true,
    plainpages=false, 
    colorlinks=true,
    linkcolor=blue,
    citecolor=blue,
    filecolor=black,
    urlcolor=blue
}
\usepackage{url}
\usepackage{verbatim}

\begin{document}
 
\title{Self-bound dipolar droplet: a localized matter-wave in free space}

\author{D.~Baillie}  
\affiliation{Department of Physics, Centre for Quantum Science,
and Dodd-Walls Centre for Photonic and Quantum Technologies, University of Otago, Dunedin, New Zealand}
 \author{R.~M.~Wilson}   
\affiliation{Department of Physics, The United States Naval Academy, Annapolis, MD 21402, USA}
\author{R.~N.~Bisset}   
\affiliation{INO-CNR BEC Center and Dipartimento di Fisica,
Universit\`a di Trento, Via Sommarive 14, I-38123 Povo, Italy}
\author{P.~B.~Blakie}   
\affiliation{Department of Physics, Centre for Quantum Science,
and Dodd-Walls Centre for Photonic and Quantum Technologies, University of Otago, Dunedin, New Zealand}

\begin{abstract}  
 We demonstrate that a dipolar condensate can be prepared into a three-dimensional wavepacket  that remains localized when released in free-space.  Such self-bound states arise from the interplay of the two-body interactions and quantum fluctuations. We develop a phase diagram for the parameter regimes where these self-bound states are stable,   examine their properties, and demonstrate how they can be produced in current experiments.
 \end{abstract} 

\maketitle

Localized structures such as solitons are of interest to a wide range of fields from photonics to many-body physics. Three-dimensionally localized light pulses, so-called light bullets, have been realized using fabricated waveguides \cite{Minardi2010a}.  The matter-wave equivalent has been the subject of numerous proposals, including using light-induced gravitational forces \cite{Giovanazzi2002a,Giovanazz2001b}, off-resonant Rydberg dressing \cite{Maucher2011a}, cold atomic gases with three-body interactions \cite{Bulgac2002a}, and spin-orbit coupled binary condensates \cite{Zhang2015a}.  However, to date none of these schemes have been realized in experiments.

\begin{figure}[t]
   \centering
  \includegraphics[width=3.4in]{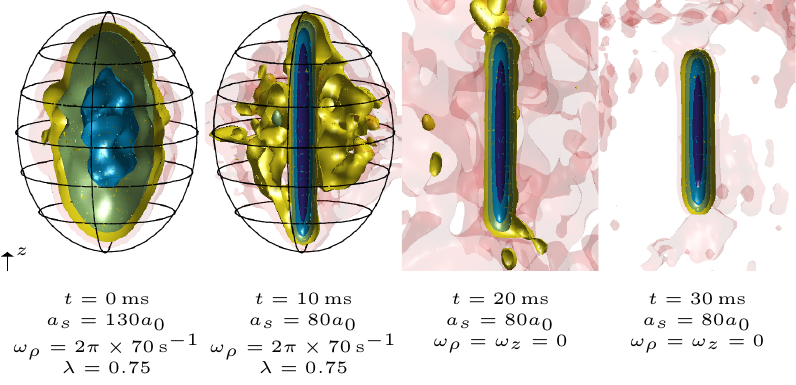} 
  \vspace*{-0.1cm}
  \caption{(color online) Density isosurfaces illustrating the dynamical production of a self-bound droplet starting from a $^{164}$Dy condensate with $a_s=130a_0$ and $10^4$ atoms. In the dynamics, $a_s$ is quenched to $80a_0$ over $10$\:ms, and then the trapping potential is turned off over $10$\:ms. 
   Contours are for a density slice in the $y=0$ plane. Each adjacent contour has a density differing by a factor of 10. See Fig.~\ref{figtime} for other simulation parameters and details.}
   \label{figiso}
\end{figure}

Here we show that it is possible to realize a localized matter wave state in current experiments with dipolar condensates  (see Fig.~\ref{figiso}). Such condensates consist of atoms with appreciable magnetic dipole moments and have been experimentally realized with chromium \cite{Griesmaier2005a,Beaufils2008}, dysprosium \cite{Mingwu2011a} and erbium \cite{Aikawa2012a}. The two-body interaction in this system includes a long-ranged and anisotropic dipole-dipole interaction (DDI) in addition to a short-ranged $s$-wave interaction \cite{Lahaye_RepProgPhys_2009}.  
For sufficiently strong dipoles the two-body interaction is partially attractive and the system is susceptible to local collapse instabilities \cite{Koch2008a,Wilson2009a,Lahaye2009a}.
However, recent experiments exploring this regime with trapped dipolar condensates have observed the formation of droplet arrays, i.e.~the atoms coalesce into a set of small and dense droplets that have long life-times ($\gtrsim\!100\,$ms)  \cite{Kadau2016a,Ferrier-Barbut2016a,Bisset2016a}.  Recent works demonstrated that quantum fluctuations are most likely responsible for stabilizing these droplets \cite{Ferrier-Barbut2016a,Wachtler2016a,Saito2016a}, as they contribute a local energy proportional to $n^{5/2}$, where $n$ is the density, that arrests the two-body driven collapse (proportional to $n^2$) at sufficiently high $n$. 

In this work we develop a general theory of self-bound dipolar condensates based on the generalized non-local Gross-Pitaevskii equation (GPE) that includes corrections due to quantum fluctuations.   We obtain self-bound states directly using numerical methods \cite{Ronen2006a} and by an approximate variational approach. This allows us to construct a phase diagram for the regime of interaction parameters and atom number $N$ where self-bound states exist, and to explore the typical properties of these states. Finally, we discuss how these states can be produced in experiments beginning from a trapped dipolar condensate by dynamically adjusting the trapping potential and $s$-wave scattering length. These results show that the lifetimes of the self-bound states are ultimately set by the three-body loss rate, which eventually reduces the atom number to the point when the wavepacket is no longer self-bound.

\emph{Formalism}-- 
The meanfield theory for the dynamics of a dipolar condensate is given by a generalized non-local GPE 
\begin{align}
i\hbar\dot{\psi}=\left[-\frac{\hbar^2\nabla^2}{2m}+g|\psi|^2+\Phi(\mathbf{r})+\gamma_{\mathrm{QF}}|\psi|^3\right]\psi(\mathbf{r}),\label{GGPE}
\end{align}
where $\psi$ is the condensate wavefunction. Here  $g=4\pi a_s\hbar^2/m$ is the $s$-wave coupling constant with $a_s$ being the $s$-wave scattering length. 
The DDIs are described by the term $\Phi(\mathbf{r})=\int d\mathbf{r}'U_{\mathrm{dd}}(\mathbf{r}-\mathbf{r}')|\psi(\mathbf{r}')|^2$, where $U_{\mathrm{dd}}(\mathbf{r})=\frac{\mu_0\mu^2}{4\pi r^3}(1-3\cos^2\theta)$ and $\theta$ is the angle between $\mathbf{r}$ and the polarization axis of the dipoles, which we take to be the $z$ direction. 
To leading order the quantum fluctuation correction to the meanfield energy for a uniform dipolar condensate is $\Delta E=\frac{2}{5}\gamma_{\mathrm{QF}}n^{5/2}$ \cite{Lima2011a}, with coefficient 
\cite{Ferrier-Barbut2016a,Bisset2016a}
\begin{equation}
\gamma_{\mathrm{QF}}=\frac{32}{3}g\sqrt{\frac{a_s^3}{\pi}}\left(1+\frac{3}{2}\epsilon_{\mathrm{dd}}^2\right).
\end{equation} 
Here $\epsilon_{\mathrm{dd}}=a_{\mathrm{dd}}/a_s$ is the ratio of DDI to $s$-wave interaction strengths and $a_{\mathrm{dd}}=m\mu_0\mu^2/12\pi\hbar^2$ is the {dipole length}  \cite{Lahaye_RepProgPhys_2009}. In Eq.~(\ref{GGPE}) these fluctuations are included via the associated chemical potential shift $\Delta\mu=\gamma_{\mathrm{QF}}n^{3/2}$ making the local density approximation $n^{3/2}\to|\psi|^3$.  The applicability of generalized GPE (\ref{GGPE}) to dipolar condensates in the regime we consider here has been discussed in Refs.~\cite{Wachtler2016a,Saito2016a,Bisset2016a}.

\begin{figure}[htbp]
   \centering
  \includegraphics[width=\columnwidth]{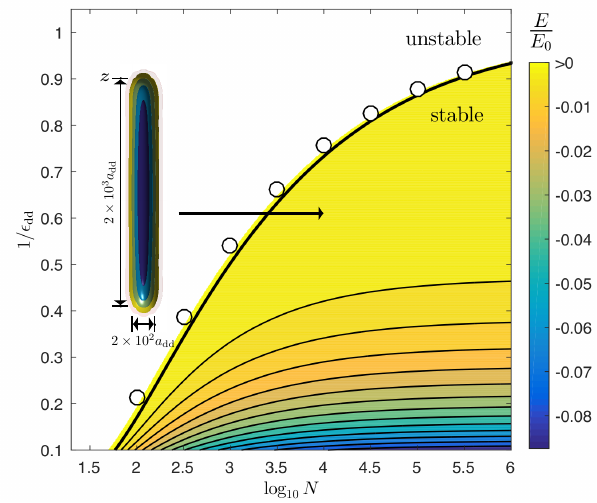} 
  \vspace*{-0.1cm}
   \caption{(color online)  Phase diagram of  self-bound solutions as a function of $1/\epsilon_{\mathrm{dd}}$ and $N$ calculated using the variational approach.   The colors show the energy of the solutions.  The thick black line corresponds to $E=0$. Colored regions above this line are metastable (i.e.~$E>0$) and in the white region only the trivial dispersed solution exists. Stability thresholds from GPE calculations are indicated by circles (also see Fig.~\ref{figDropProps}).
   The inset shows isodensity contours of a self-bound GPE (parameters indicated by arrow). The contours have the same scale as Fig.~\ref{figiso}.  
}
   \label{figPD}
\end{figure} 

Equation~(\ref{GGPE}) possesses a continuous translational symmetry (due to the absence of an external trapping potential) and has a trivial uniform stationary state $\psi=\sqrt{n}$ with zero energy  (noting that $n\to0$   for fixed  $N$). Here our interest is in localized (self-bound) stationary solutions to Eq.~(\ref{GGPE}). If these localized solutions have negative energy then they are thermodynamically stable with respect to the trivial uniform solution. 

A useful description of the system is furnished by a Gaussian variational ansatz for the condensate wavefunction 
\[\psi_{\mathrm{v}}(\mathbf{r})=\sqrt{\tfrac{8N}{\pi^{3/2}\sigma_\rho^2\sigma_z}}e^{-2(\rho^2/\sigma_\rho^2+z^2/\sigma_z^2)},\]  where $\sigma_\rho$ and $\sigma_z$ are the variational width parameters and we have utilized the cylindrical symmetry of the system around the $z$ axis. The equilibrium width parameters can be determined by finding minima of the energy functional associated with Eq.~(\ref{GGPE}), which has the form \cite{Bisset2016a}
\begin{align}
\frac{E(\sigma_\rho,\sigma_z)}{E_0/N}&=\frac{2}{\bar{\sigma}^2_\rho}+\frac{1}{\bar{\sigma}^2_z}+\frac{8}{\sqrt{2\pi}\bar{\sigma}_\rho^2\bar{\sigma}_z}\left[\frac{1}{\epsilon_{\mathrm{dd}}}-f\left(\frac{\bar{\sigma}_\rho}{\bar{\sigma}_z}\right)\right]\nonumber \\
 &+c\frac{1+\frac{3}{2}\epsilon_{\mathrm{dd}}^2}{\bar{\sigma}_\rho^3\bar{\sigma}_z^{3/2}N\epsilon_{\mathrm{dd}}^{5/2}},\label{Evar}
\end{align}
where $c=2^{14}/75\sqrt{5}\pi^{7/4}\approx13.18$ and $
 f(x)=\frac{1+2x^2}{1-x^2}-\frac{3x^2\mathrm{arctanh}\sqrt{1-x^2}}{(1-x^2)^{3/2}}$. Here  $E_0=\hbar^2/ma_{\mathrm{dd}}^2$  and $L_0=Na_{\mathrm{dd}}$ are convenient units of energy and length, and we define $\bar{\sigma}_\nu\equiv\sigma_\nu/L_0$. 
 
\emph{Equilibrium results}-- 
 The form of Eq.~(\ref{Evar}) reveals that solution properties only depend on the parameters $N$ and $1/\epsilon_{\mathrm{dd}}$. In terms of these parameters the energy of non-trivial  solutions [i.e.~local minima to (\ref{Evar})] are shown in Fig.~\ref{figPD} as a ``phase diagram'' for the existence of these solutions.  We mark the phase boundary (the binodal line, where $E=0$) with a thick black line, below which  these solutions are the stable ground state. We note that localized solutions persist slightly beyond this region as meta-stable  states, until they reach the spinodal line.

\begin{figure}[tbp]
   \centering
  \includegraphics[width=\columnwidth]{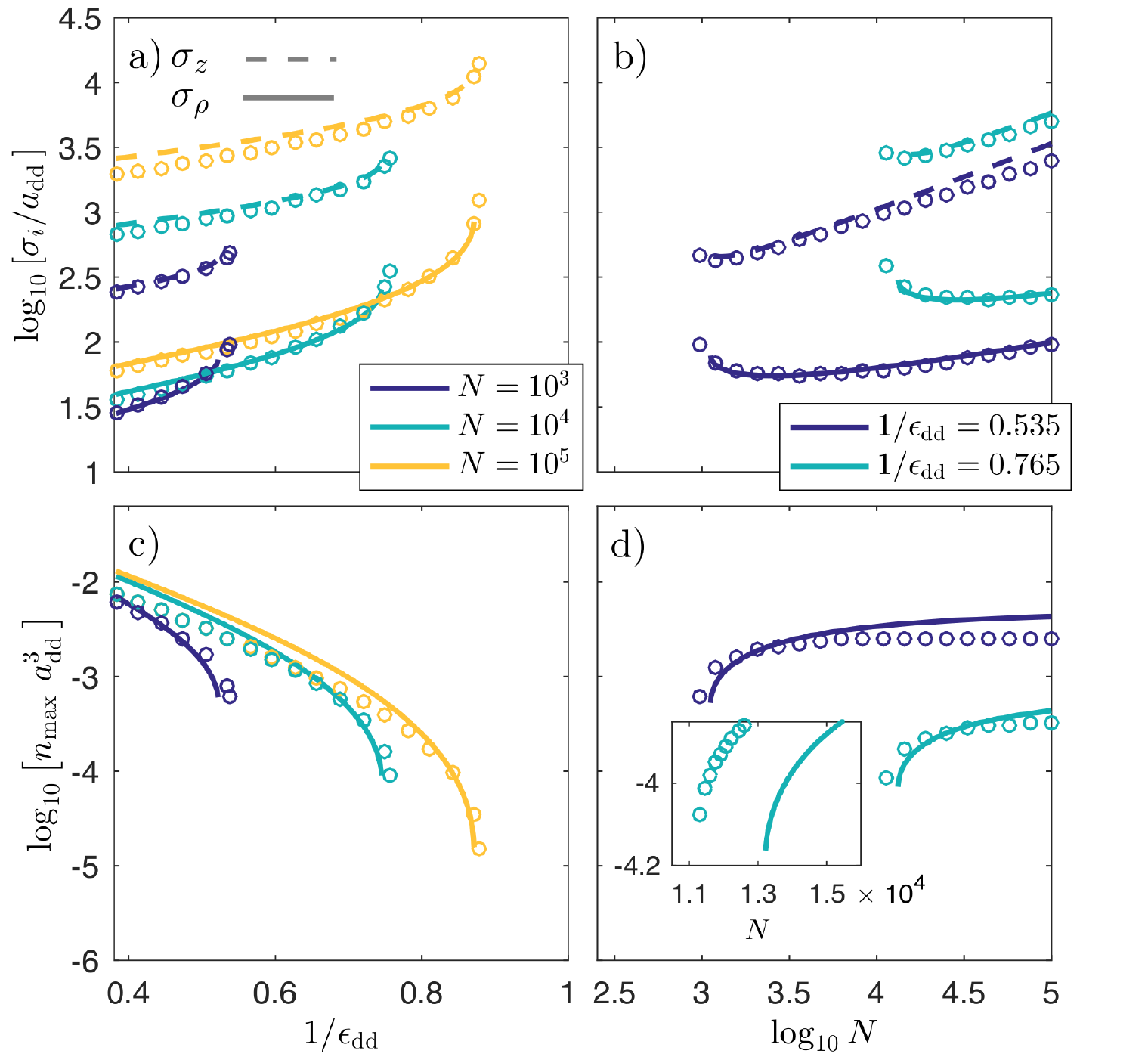} 
  \vspace*{-0.1cm}
   \caption{(color online)  
   Properties of self-bound solutions. Widths $\{\sigma_\rho,\sigma_z\}$ as a function of (a) $1/\epsilon_{\mathrm{dd}}$ for $N=10^3$ (black), $10^4$ (gray), and $10^5$ (light gray) and (b) $N$ for $1/\epsilon_{\mathrm{dd}} = 0.535$ (black) and $0.765$ (gray). Peak number density as a function of (c) $1/\epsilon_{\mathrm{dd}}$ and (d) $N$. Variational predictions are shown as solid and dashed lines, and GPE results are indicated as circles. The inset to (d) shows the tip of the $\epsilon_{\mathrm{dd}}^{-1}=0.765$ results with a linear horizontal axis to better reveal the difference in variational and GPE predictions for the minimum atom number. The two choices of $\epsilon_{\mathrm{dd}}$ in (b) and (d) correspond to  $a_s = 70.0$ and $100 a_0$ for Dy, and $a_s = 35.6$ and $50.9 a_0$ for Er.  
      }
   \label{figDropProps}
\end{figure} 

 The maximum value of $\epsilon_{\mathrm{dd}}^{-1}$ at which the localized state becomes unstable increases with $N$, approaching unity as $N$ gets large. Equivalently,  for any given value of $\epsilon_{\mathrm{dd}}^{-1}<1$ there will be a minimum value of $N$ below which the localized solution becomes unstable. We show later that this minimum number has important implications when atomic loss is accounted for and can limit the lifetime of the self-bound state.

Other properties of the droplet solutions are considered in Fig.~\ref{figDropProps}.  Here we show results from both variational solutions and numerically determined stationary solutions of the generalized GPE (\ref{GGPE}). Figure \ref{figDropProps}(a) shows that the self-bound solutions are elongated along $z$ (i.e.~$\sigma_z\gg\sigma_\rho$) since in this configuration the DDI-term [i.e.~$-f$ term  in Eq.~(\ref{Evar})] becomes negative, arising from the dominant attractive head-to-tail interaction between dipoles.  As $\epsilon_{\mathrm{dd}}^{-1}$ increases the widths monotonically increase until the spinodal stability threshold is reached, where they diverge.  Considering the widths as a function of $N$ in Fig.~\ref{figDropProps}(b) reveals that the widths monotonically increase with $N$ for sufficiently large $N$. However, as $N$ approaches the minimum value (where the lines terminate),  the widths start increasing with decreasing $N$. This occurs because at low $N$ the kinetic energy (quantum pressure) becomes important and can destabilise the self-bound state. 

Figure~\ref{figDropProps}(c)  shows that the maximum density $n_{\max}$ decreases with increasing $\epsilon_{\mathrm{dd}}^{-1}$. As $\epsilon_{\mathrm{dd}}^{-1}$ increases, the  two-body interactions become less attractive and the quantum fluctuation term is able to balance their effect at low $n$. This result also shows that the value of the diluteness parameter $na_{\mathrm{dd}}^3$ remains $\lesssim10^{-2}$ over the parameters considered here.
Considering $n_{\max}$ as a function of $N$ in Fig.~~\ref{figDropProps}(d)  reveals that except near the minimum number, the droplet density saturates, i.e.~adding more particles to the system barely changes  $n_{\max}$, but instead causes the droplet to expand. This behaviour is reminiscent of liquid states of matter.

\begin{figure}[tbp]
   \centering
  \includegraphics[width=\columnwidth]{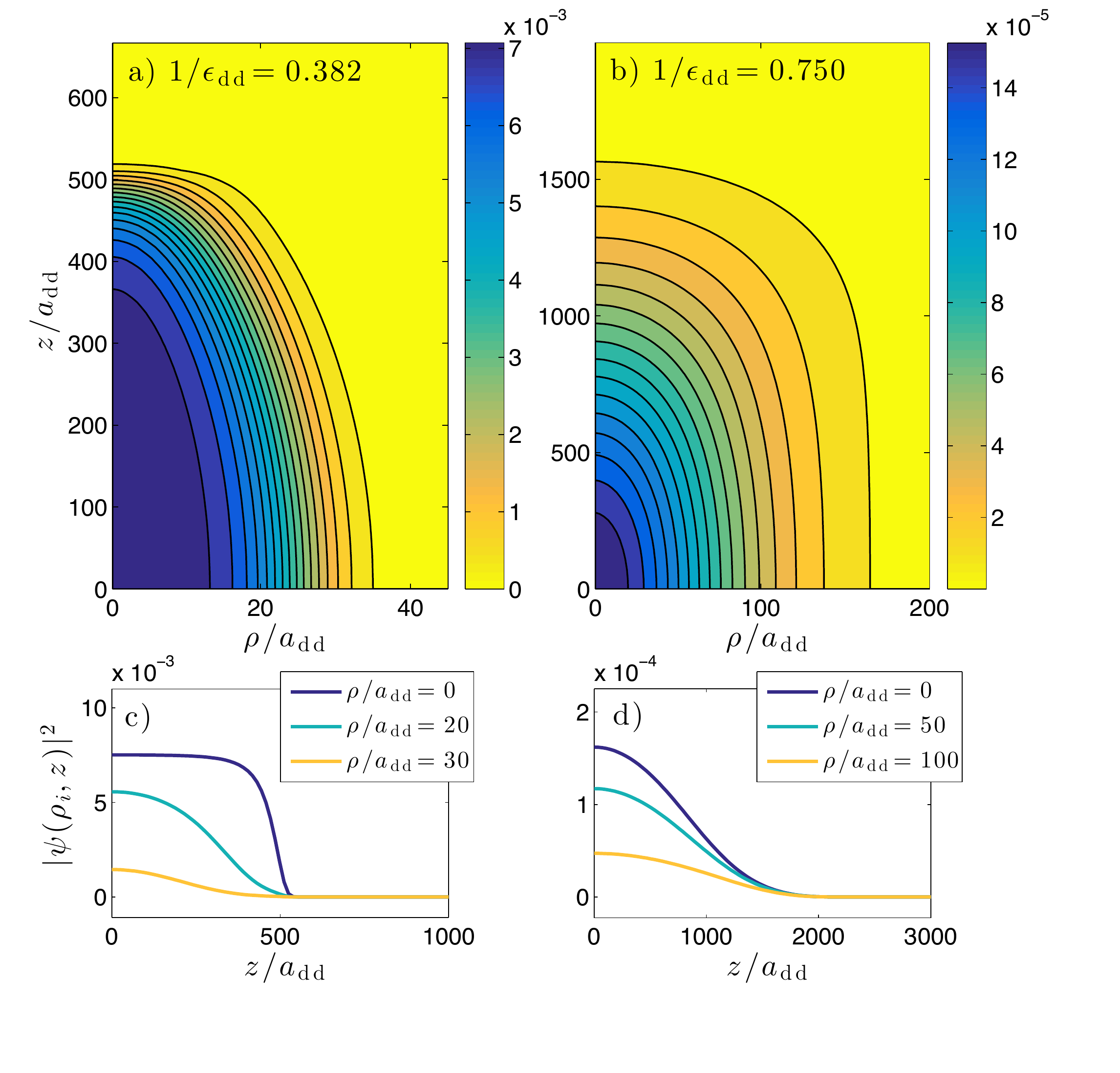} 
  \vspace*{-0.1cm}
   \caption{(color online)  
   Droplet density profiles for $N=10^4$.   (a)  Self-bound solution  with $E=-5.2\times10^{-3}E_0$.  (b) Solution near the instability threshold with $E=0$.  (c)  Axial density slices of the self-bound solution shown in (a) for $\rho=0$ (black), $\rho=20a_\mathrm{dd}$ (gray) and $\rho=30a_\mathrm{dd}$ (light gray).  (d) Axial density slices of the solution shown in (b) for $\rho=0$ (black), $\rho=50a_\mathrm{dd}$ (gray) and $\rho=100a_\mathrm{dd}$ (light gray). Note the different scales on the axes for the two results.   The values of $1/\epsilon_{\mathrm{dd}}$ are chosen to  correspond to $a_s = 50.0$ and $98.0 a_0$ for Dy, and $a_s = 25.4$ and $49.9 a_0$ for Er.
    }
   \label{figGpeWfns}
\end{figure}

Examples of GPE solutions are shown in Fig.~\ref{figGpeWfns}. Notably, the self-bound solution shown in Fig.~\ref{figGpeWfns}(a) is more deeply bound, i.e.~is in a regime where the maximum density has saturated as a function of $N$. As a result this droplet has a flat density profile along the $z$ axis [see Fig.~\ref{figGpeWfns}(c)]. In contrast Fig.~\ref{figGpeWfns}(b) shows a self-bound solution with $E=0$, at the threshold of metastability. This state is much larger, has a significantly lower peak density, and does not exhibit density saturation effects [see Fig.~\ref{figGpeWfns}(d)].  
    
   \begin{figure}[htbp]
   \centering
  \includegraphics[width=\columnwidth]{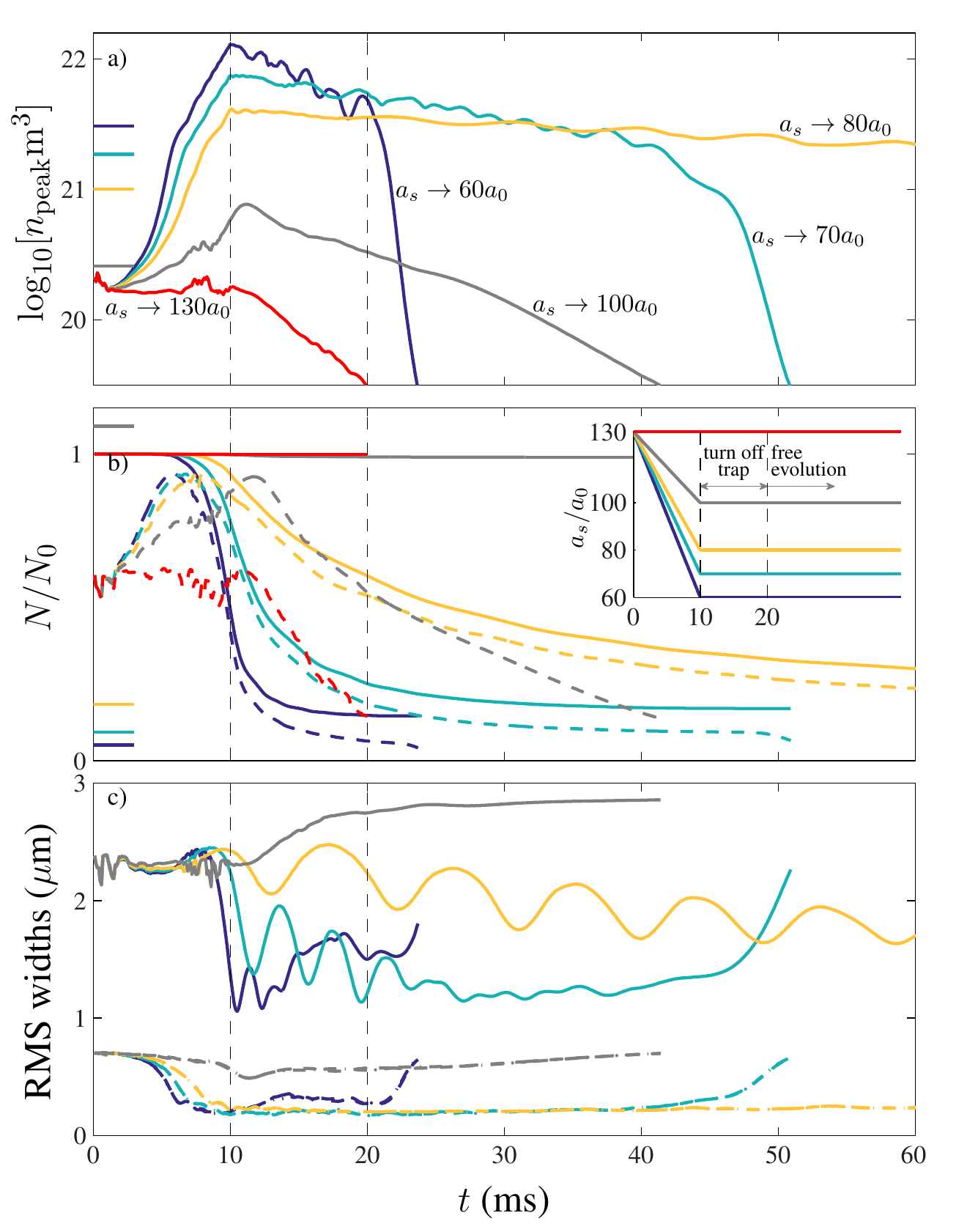} 
  \vspace*{-0.25cm}
  \caption{(color online) Droplet formation for $^{164}$Dy starting with $N_0=10^4$ atoms (plus $T\!=\!20$\:nK initial noise) in a harmonic trapping potential with $\omega_\rho=2\pi\times 70$\:Hz and $\lambda = 0.75$. Different line colors indicate the final $a_s$ value [see inset to (b)]. (a) peak density, colored ticks by the vertical axis indicate the equilibrium critical peak density, (b) total number (solid) and number within the cylinder of diameter 
  $3\:\mu$m and height $10\:\mu$m centered at the location of $n_{\mathrm{peak}}$ (dashed), colored ticks indicate the equilibrium critical droplet number from Fig.~\ref{figPD},  (c) oscillation of the RMS widths $\sqrt{\langle r_i^2\rangle -\langle r_i \rangle^2}$ for $r_i=z$ (solid), $r_i=x$ (dashed) and $r_i=y$ (dash-dotted) evaluated using the field within the  cylinder.  The inset to (b) shows the  time sequence used: linear $a_s$ quench to $[60,70,80,100,130]a_0$ over $\tau_{s}=10$\:ms, followed by a $\tau_{\mathrm{trap}}=10$\:ms linear ramp of the trap frequencies to zero.
  For $t\ge20$\:ms the system evolves in free space.}
   \label{figtime}
\end{figure} 

\emph{Dynamical Results}-- 
 We now turn to considering how a self-bound droplet can be obtained from the type of trapped condensate typically prepared in experiments.
 From this initial state we propose using a sequence of two ramps [see inset to Fig.~\ref{figtime}(b)] to produce the desired state in free space: (i) the $s$-wave scattering length is reduced over a time scale $\tau_{s}$ using a Feshbach resonance until $\epsilon_{\mathrm{dd}}^{-1}$ reaches a value necessary for a self-bound droplet; (ii) the trapping potential is then ramped off over a time scale $\tau_{\mathrm{trap}}$ leaving the droplet in a self-bound state.
 
To accurately model the dynamics of formation it is necessary to augment the generalized GPE (\ref{GGPE}) with the additional terms $\mathcal{V}\psi$ where 
\begin{equation}
\mathcal{V} \equiv\tfrac{1}{2}m\omega_\rho(t)^2(\rho^2+\lambda^2z^2)-\tfrac{i}{2}{\hbar L_3}|\psi|^4,
\end{equation}
that describe the harmonic trapping potential and three-body loss processes that will occur at high atomic density. Here $\omega_\rho(t)$ is the radial trap frequency at time $t$,  $\lambda=\omega_z/\omega_\rho$ is the trap aspect ratio, and $L_3$ is the loss coefficient. 
We simulate the GPE dynamics using a 3D Fourier method on a grid of $512\times512\times256$ points evolved using a fourth-order Runge-Kutta algorithm.
The initial condition for the dynamics, $\psi_0$, is a stationary solution of the GPE subject to the trapping potential with $\epsilon_{\mathrm{dd}}\approx1$ and noise added to mimic quantum and thermal fluctuations. The procedure  for adding this noise is as described in Refs.~\cite{Bisset2015a,Blakie2016a}.
 
We choose to use a  prolate trap ($\lambda<1$) since the condensate continuously evolves into a single droplet state  as $a_s$ (i.e.~$\epsilon_{\mathrm{dd}}^{-1}$) is reduced   \cite{Bisset2016a}. In contrast, for oblate traps there is a first order phase transition between the stable condensate and droplet state, and the heating that occurs when crossing this transition leads to multiple droplets forming (also see \cite{Blakie2016a,Wachtler2016a}).
For these droplets to remain self-bound as the trap is reduced each must exceed the minimum atom number for stability. Their mutual interaction will, however, cause them to repel and move away from each other.

The results of selected dynamical simulations are summarized in Fig.~\ref{figtime} for parameters relevant to $^{164}$Dy with $L_3 = 1.2\times 10^{-41}\:\mathrm{m}^6\mathrm{s}^{-1}$ \cite{Wachtler2016a}. The initial state $\psi_0$ uses $a_s=130a_0$. The peak density is seen to rapidly increase as $a_s$ is reduced, signaling the droplet formation.
Once fully formed, the peak density of the self-bound solution is higher if $a_s$ is quenched to a lower value [also see Fig.~\ref{figDropProps}(c)], and as a result the atom number decreases through three-body loss most rapidly for smaller $a_s$.
In order to distinguish non-self-bound atoms (which are expelled as the trap is turned off) from those in the droplet we evaluate both the total atom number and the number within a cylindrical region centered on the droplet in Fig.~\ref{figtime}(b). The similar behavior of the decay in total number and droplet number shows that most of the atom loss occurs in the dense droplet, and shows that the droplet atoms remain localized in the cylindrical region. We also observe in the two lowest ($a_s$) quenches considered that the steady decay in droplet number is suddenly interrupted by a more rapid decay at later times. This occurs because when the atom number falls below the minimum number for a stable droplet [which depends on $a_s$, indicated by the colored ticks by the vertical axis in Fig.~\ref{figtime}], the droplet suddenly becomes unbounded and disperses. For the quench to $a_s=80a_0$ this does not occur within the time range we simulate (also see Fig.~\ref{figiso}).
For the  quench to $a_s=100a_0$ the droplet does not form at all, as can be seen by the small peak density. This is because the initial condensate number ($N_0=10^4$) is lower than the stability threshold ($1.1\times10^4$) for $a_s = 100 a_0$.

We have also investigated droplet formation sensitivity to the three-body loss rate. For $a_s=60a_0$  ($a_s=70a_0$) the droplet lifetimes are $\sim 35,20,15\:$ms  ($\sim 80,45,25\:$ms) if we scale the three-body loss parameter by the factors $\tfrac12,1,2$, respectively.

Because the quenches are reasonably fast they excite collective modes of the droplet. These excitations give rise to width oscillations that are seen to persist when the droplet is in free space, providing an additional signature of the self-bound character of the droplet.

\emph{Conclusions}-- In this paper, we have shown that dipolar condensates provide an ideal system for realizing a self-bound matterwave.  We have presented a universal description of the self-bound states  parameterized in terms of the interaction parameter ratio and atom number, thus our results are relevant for, and can be easily extended to describe experiments with Er.  Importantly, we observe that in the strongly dipolar regime droplets require a minimum atom number to be stable. We have proposed and simulated a scheme for producing a self-bound droplet in free space in a parameter regime accessible to current experiments. We show that three-body loss plays an important role and will ultimately limit self-bound droplet lifetime. However, our results show that quenching to larger values of $\epsilon_{\mathrm{dd}}^{-1}$  are favourable for producing a long lived droplet because of the slower loss rates.\\

\emph{Acknowledgments}-- 
The authors acknowledge valuable conversations with F.~Ferlaino. 
DB and PBB acknowledge the contribution of NZ eScience Infrastructure (NeSI) high-performance computing facilities, and support from the Marsden Fund of the Royal Society of New Zealand.   
RMW acknowledges partial support from the National Science Foundation under Grant No.~PHYS-1516421. 
 RNB acknowledges support by the QUIC grant of the Horizon2020 FET program and by Provincia Autonoma di Trento.\\

\emph{Note added}-- In the final stages of  preparing of this manuscript we became aware of Ref.~\cite{Wachtler2016c}, which discusses  equilibrium properties of trapped and self-bound droplets.


\begin{thebibliography}{24}%
\makeatletter
\providecommand \@ifxundefined [1]{%
 \@ifx{#1\undefined}
}%
\providecommand \@ifnum [1]{%
 \ifnum #1\expandafter \@firstoftwo
 \else \expandafter \@secondoftwo
 \fi
}%
\providecommand \@ifx [1]{%
 \ifx #1\expandafter \@firstoftwo
 \else \expandafter \@secondoftwo
 \fi
}%
\providecommand \natexlab [1]{#1}%
\providecommand \enquote  [1]{``#1''}%
\providecommand \bibnamefont  [1]{#1}%
\providecommand \bibfnamefont [1]{#1}%
\providecommand \citenamefont [1]{#1}%
\providecommand \href@noop [0]{\@secondoftwo}%
\providecommand \href [0]{\begingroup \@sanitize@url \@href}%
\providecommand \@href[1]{\@@startlink{#1}\@@href}%
\providecommand \@@href[1]{\endgroup#1\@@endlink}%
\providecommand \@sanitize@url [0]{\catcode `\\12\catcode `\$12\catcode
  `\&12\catcode `\#12\catcode `\^12\catcode `\_12\catcode `\%12\relax}%
\providecommand \@@startlink[1]{}%
\providecommand \@@endlink[0]{}%
\providecommand \url  [0]{\begingroup\@sanitize@url \@url }%
\providecommand \@url [1]{\endgroup\@href {#1}{\urlprefix }}%
\providecommand \urlprefix  [0]{URL }%
\providecommand \Eprint [0]{\href }%
\providecommand \doibase [0]{http://dx.doi.org/}%
\providecommand \selectlanguage [0]{\@gobble}%
\providecommand \bibinfo  [0]{\@secondoftwo}%
\providecommand \bibfield  [0]{\@secondoftwo}%
\providecommand \translation [1]{[#1]}%
\providecommand \BibitemOpen [0]{}%
\providecommand \bibitemStop [0]{}%
\providecommand \bibitemNoStop [0]{.\EOS\space}%
\providecommand \EOS [0]{\spacefactor3000\relax}%
\providecommand \BibitemShut  [1]{\csname bibitem#1\endcsname}%
\let\auto@bib@innerbib\@empty
\bibitem [{\citenamefont {Minardi}\ \emph {et~al.}(2010)\citenamefont
  {Minardi}, \citenamefont {Eilenberger}, \citenamefont {Kartashov},
  \citenamefont {Szameit}, \citenamefont {R\"opke}, \citenamefont {Kobelke},
  \citenamefont {Schuster}, \citenamefont {Bartelt}, \citenamefont {Nolte},
  \citenamefont {Torner}, \citenamefont {Lederer}, \citenamefont
  {T\"unnermann},\ and\ \citenamefont {Pertsch}}]{Minardi2010a}%
  \BibitemOpen
  \bibfield  {author} {\bibinfo {author} {\bibfnamefont {S.}~\bibnamefont
  {Minardi}}, \bibinfo {author} {\bibfnamefont {F.}~\bibnamefont
  {Eilenberger}}, \bibinfo {author} {\bibfnamefont {Y.~V.}\ \bibnamefont
  {Kartashov}}, \bibinfo {author} {\bibfnamefont {A.}~\bibnamefont {Szameit}},
  \bibinfo {author} {\bibfnamefont {U.}~\bibnamefont {R\"opke}}, \bibinfo
  {author} {\bibfnamefont {J.}~\bibnamefont {Kobelke}}, \bibinfo {author}
  {\bibfnamefont {K.}~\bibnamefont {Schuster}}, \bibinfo {author}
  {\bibfnamefont {H.}~\bibnamefont {Bartelt}}, \bibinfo {author} {\bibfnamefont
  {S.}~\bibnamefont {Nolte}}, \bibinfo {author} {\bibfnamefont
  {L.}~\bibnamefont {Torner}}, \bibinfo {author} {\bibfnamefont
  {F.}~\bibnamefont {Lederer}}, \bibinfo {author} {\bibfnamefont
  {A.}~\bibnamefont {T\"unnermann}}, \ and\ \bibinfo {author} {\bibfnamefont
  {T.}~\bibnamefont {Pertsch}},\ }\href {\doibase %
  10.1103/PhysRevLett.105.263901} {\bibfield  {journal} {\bibinfo  {journal}
  {Phys. Rev. Lett.}\ }\textbf {\bibinfo {volume} {105}},\ \bibinfo {pages}
  {263901} (\bibinfo {year} {2010})}\BibitemShut {NoStop}%
\bibitem [{\citenamefont {Giovanazzi}\ \emph {et~al.}(2002)\citenamefont
  {Giovanazzi}, \citenamefont {O'Dell},\ and\ \citenamefont
  {Kurizki}}]{Giovanazzi2002a}%
  \BibitemOpen
  \bibfield  {author} {\bibinfo {author} {\bibfnamefont {S.}~\bibnamefont
  {Giovanazzi}}, \bibinfo {author} {\bibfnamefont {D.}~\bibnamefont {O'Dell}},
  \ and\ \bibinfo {author} {\bibfnamefont {G.}~\bibnamefont {Kurizki}},\ }\href
  {\doibase 10.1103/PhysRevLett.88.130402} {\bibfield  {journal} {\bibinfo
  {journal} {Phys. Rev. Lett.}\ }\textbf {\bibinfo {volume} {88}},\ \bibinfo
  {pages} {130402} (\bibinfo {year} {2002})}\BibitemShut {NoStop}%
\bibitem [{\citenamefont {Giovanazzi}\ \emph {et~al.}(2001)\citenamefont
  {Giovanazzi}, \citenamefont {O'Dell},\ and\ \citenamefont
  {Kurizki}}]{Giovanazz2001b}%
  \BibitemOpen
  \bibfield  {author} {\bibinfo {author} {\bibfnamefont {S.}~\bibnamefont
  {Giovanazzi}}, \bibinfo {author} {\bibfnamefont {D.}~\bibnamefont {O'Dell}},
  \ and\ \bibinfo {author} {\bibfnamefont {G.}~\bibnamefont {Kurizki}},\ }\href
  {\doibase 10.1103/PhysRevA.63.031603} {\bibfield  {journal} {\bibinfo
  {journal} {Phys. Rev. A}\ }\textbf {\bibinfo {volume} {63}},\ \bibinfo
  {pages} {031603} (\bibinfo {year} {2001})}\BibitemShut {NoStop}%
\bibitem [{\citenamefont {Maucher}\ \emph {et~al.}(2011)\citenamefont
  {Maucher}, \citenamefont {Henkel}, \citenamefont {Saffman}, \citenamefont
  {Kr\'olikowski}, \citenamefont {Skupin},\ and\ \citenamefont
  {Pohl}}]{Maucher2011a}%
  \BibitemOpen
  \bibfield  {author} {\bibinfo {author} {\bibfnamefont {F.}~\bibnamefont
  {Maucher}}, \bibinfo {author} {\bibfnamefont {N.}~\bibnamefont {Henkel}},
  \bibinfo {author} {\bibfnamefont {M.}~\bibnamefont {Saffman}}, \bibinfo
  {author} {\bibfnamefont {W.}~\bibnamefont {Kr\'olikowski}}, \bibinfo {author}
  {\bibfnamefont {S.}~\bibnamefont {Skupin}}, \ and\ \bibinfo {author}
  {\bibfnamefont {T.}~\bibnamefont {Pohl}},\ }\href {\doibase %
  10.1103/PhysRevLett.106.170401} {\bibfield  {journal} {\bibinfo  {journal}
  {Phys. Rev. Lett.}\ }\textbf {\bibinfo {volume} {106}},\ \bibinfo {pages}
  {170401} (\bibinfo {year} {2011})}\BibitemShut {NoStop}%
\bibitem [{\citenamefont {Bulgac}(2002)}]{Bulgac2002a}%
  \BibitemOpen
  \bibfield  {author} {\bibinfo {author} {\bibfnamefont {A.}~\bibnamefont
  {Bulgac}},\ }\href {\doibase 10.1103/PhysRevLett.89.050402} {\bibfield
  {journal} {\bibinfo  {journal} {Phys. Rev. Lett.}\ }\textbf {\bibinfo
  {volume} {89}},\ \bibinfo {pages} {050402} (\bibinfo {year}
  {2002})}\BibitemShut {NoStop}%
\bibitem [{\citenamefont {Zhang}\ \emph {et~al.}(2015)\citenamefont {Zhang},
  \citenamefont {Zhou}, \citenamefont {Malomed},\ and\ \citenamefont
  {Pu}}]{Zhang2015a}%
  \BibitemOpen
  \bibfield  {author} {\bibinfo {author} {\bibfnamefont {Y.-C.}\ \bibnamefont
  {Zhang}}, \bibinfo {author} {\bibfnamefont {Z.-W.}\ \bibnamefont {Zhou}},
  \bibinfo {author} {\bibfnamefont {B.~A.}\ \bibnamefont {Malomed}}, \ and\
  \bibinfo {author} {\bibfnamefont {H.}~\bibnamefont {Pu}},\ }\href {\doibase %
  10.1103/PhysRevLett.115.253902} {\bibfield  {journal} {\bibinfo  {journal}
  {Phys. Rev. Lett.}\ }\textbf {\bibinfo {volume} {115}},\ \bibinfo {pages}
  {253902} (\bibinfo {year} {2015})}\BibitemShut {NoStop}%
\bibitem [{\citenamefont {Griesmaier}\ \emph {et~al.}(2005)\citenamefont
  {Griesmaier}, \citenamefont {Werner}, \citenamefont {Hensler}, \citenamefont
  {Stuhler},\ and\ \citenamefont {Pfau}}]{Griesmaier2005a}%
  \BibitemOpen
  \bibfield  {author} {\bibinfo {author} {\bibfnamefont {A.}~\bibnamefont
  {Griesmaier}}, \bibinfo {author} {\bibfnamefont {J.}~\bibnamefont {Werner}},
  \bibinfo {author} {\bibfnamefont {S.}~\bibnamefont {Hensler}}, \bibinfo
  {author} {\bibfnamefont {J.}~\bibnamefont {Stuhler}}, \ and\ \bibinfo
  {author} {\bibfnamefont {T.}~\bibnamefont {Pfau}},\ }\href {\doibase %
  10.1103/PhysRevLett.94.160401} {\bibfield  {journal} {\bibinfo  {journal}
  {Phys. Rev. Lett.}\ }\textbf {\bibinfo {volume} {94}},\ \bibinfo {pages}
  {160401} (\bibinfo {year} {2005})}\BibitemShut {NoStop}%
\bibitem [{\citenamefont {Beaufils}\ \emph {et~al.}(2008)\citenamefont
  {Beaufils}, \citenamefont {Chicireanu}, \citenamefont {Zanon}, \citenamefont
  {Laburthe-Tolra}, \citenamefont {Mar\'echal}, \citenamefont {Vernac},
  \citenamefont {Keller},\ and\ \citenamefont {Gorceix}}]{Beaufils2008}%
  \BibitemOpen
  \bibfield  {author} {\bibinfo {author} {\bibfnamefont {Q.}~\bibnamefont
  {Beaufils}}, \bibinfo {author} {\bibfnamefont {R.}~\bibnamefont
  {Chicireanu}}, \bibinfo {author} {\bibfnamefont {T.}~\bibnamefont {Zanon}},
  \bibinfo {author} {\bibfnamefont {B.}~\bibnamefont {Laburthe-Tolra}},
  \bibinfo {author} {\bibfnamefont {E.}~\bibnamefont {Mar\'echal}}, \bibinfo
  {author} {\bibfnamefont {L.}~\bibnamefont {Vernac}}, \bibinfo {author}
  {\bibfnamefont {J.-C.}\ \bibnamefont {Keller}}, \ and\ \bibinfo {author}
  {\bibfnamefont {O.}~\bibnamefont {Gorceix}},\ }\href {\doibase %
  10.1103/PhysRevA.77.061601} {\bibfield  {journal} {\bibinfo  {journal} {Phys.
  Rev. A}\ }\textbf {\bibinfo {volume} {77}},\ \bibinfo {pages} {061601}
  (\bibinfo {year} {2008})}\BibitemShut {NoStop}%
\bibitem [{\citenamefont {Lu}\ \emph {et~al.}(2011)\citenamefont {Lu},
  \citenamefont {Burdick}, \citenamefont {Youn},\ and\ \citenamefont
  {Lev}}]{Mingwu2011a}%
  \BibitemOpen
  \bibfield  {author} {\bibinfo {author} {\bibfnamefont {M.}~\bibnamefont
  {Lu}}, \bibinfo {author} {\bibfnamefont {N.~Q.}\ \bibnamefont {Burdick}},
  \bibinfo {author} {\bibfnamefont {S.~H.}\ \bibnamefont {Youn}}, \ and\
  \bibinfo {author} {\bibfnamefont {B.~L.}\ \bibnamefont {Lev}},\ }\href
  {\doibase 10.1103/PhysRevLett.107.190401} {\bibfield  {journal} {\bibinfo
  {journal} {Phys. Rev. Lett.}\ }\textbf {\bibinfo {volume} {107}},\ \bibinfo
  {pages} {190401} (\bibinfo {year} {2011})}\BibitemShut {NoStop}%
\bibitem [{\citenamefont {Aikawa}\ \emph {et~al.}(2012)\citenamefont {Aikawa},
  \citenamefont {Frisch}, \citenamefont {Mark}, \citenamefont {Baier},
  \citenamefont {Rietzler}, \citenamefont {Grimm},\ and\ \citenamefont
  {Ferlaino}}]{Aikawa2012a}%
  \BibitemOpen
  \bibfield  {author} {\bibinfo {author} {\bibfnamefont {K.}~\bibnamefont
  {Aikawa}}, \bibinfo {author} {\bibfnamefont {A.}~\bibnamefont {Frisch}},
  \bibinfo {author} {\bibfnamefont {M.}~\bibnamefont {Mark}}, \bibinfo {author}
  {\bibfnamefont {S.}~\bibnamefont {Baier}}, \bibinfo {author} {\bibfnamefont
  {A.}~\bibnamefont {Rietzler}}, \bibinfo {author} {\bibfnamefont
  {R.}~\bibnamefont {Grimm}}, \ and\ \bibinfo {author} {\bibfnamefont
  {F.}~\bibnamefont {Ferlaino}},\ }\href {\doibase %
  10.1103/PhysRevLett.108.210401} {\bibfield  {journal} {\bibinfo  {journal}
  {Phys. Rev. Lett.}\ }\textbf {\bibinfo {volume} {108}},\ \bibinfo {pages}
  {210401} (\bibinfo {year} {2012})}\BibitemShut {NoStop}%
\bibitem [{\citenamefont {Lahaye}\ \emph {et~al.}(2009)\citenamefont {Lahaye},
  \citenamefont {Menotti}, \citenamefont {Santos}, \citenamefont {Lewenstein},\
  and\ \citenamefont {Pfau}}]{Lahaye_RepProgPhys_2009}%
  \BibitemOpen
  \bibfield  {author} {\bibinfo {author} {\bibfnamefont {T.}~\bibnamefont
  {Lahaye}}, \bibinfo {author} {\bibfnamefont {C.}~\bibnamefont {Menotti}},
  \bibinfo {author} {\bibfnamefont {L.}~\bibnamefont {Santos}}, \bibinfo
  {author} {\bibfnamefont {M.}~\bibnamefont {Lewenstein}}, \ and\ \bibinfo
  {author} {\bibfnamefont {T.}~\bibnamefont {Pfau}},\ }\href
  {http://stacks.iop.org/0034-4885/72/126401} {\bibfield  {journal} {\bibinfo
  {journal} {Rep. Prog. Phys.}\ }\textbf {\bibinfo {volume} {72}},\ \bibinfo
  {pages} {126401} (\bibinfo {year} {2009})}\BibitemShut {NoStop}%
\bibitem [{\citenamefont {Koch}\ \emph {et~al.}(2008)\citenamefont {Koch},
  \citenamefont {Lahaye}, \citenamefont {Metz}, \citenamefont {Frohlich},
  \citenamefont {Griesmaier},\ and\ \citenamefont {Pfau}}]{Koch2008a}%
  \BibitemOpen
  \bibfield  {author} {\bibinfo {author} {\bibfnamefont {T.}~\bibnamefont
  {Koch}}, \bibinfo {author} {\bibfnamefont {T.}~\bibnamefont {Lahaye}},
  \bibinfo {author} {\bibfnamefont {J.}~\bibnamefont {Metz}}, \bibinfo {author}
  {\bibfnamefont {B.}~\bibnamefont {Frohlich}}, \bibinfo {author}
  {\bibfnamefont {A.}~\bibnamefont {Griesmaier}}, \ and\ \bibinfo {author}
  {\bibfnamefont {T.}~\bibnamefont {Pfau}},\ }\href {\doibase 10.1038/nphys887}
  {\bibfield  {journal} {\bibinfo  {journal} {Nat. Phys.}\ }\textbf {\bibinfo
  {volume} {4}},\ \bibinfo {pages} {218} (\bibinfo {year} {2008})}\BibitemShut
  {NoStop}%
\bibitem [{\citenamefont {Wilson}\ \emph {et~al.}(2009)\citenamefont {Wilson},
  \citenamefont {Ronen},\ and\ \citenamefont {Bohn}}]{Wilson2009a}%
  \BibitemOpen
  \bibfield  {author} {\bibinfo {author} {\bibfnamefont {R.~M.}\ \bibnamefont
  {Wilson}}, \bibinfo {author} {\bibfnamefont {S.}~\bibnamefont {Ronen}}, \
  and\ \bibinfo {author} {\bibfnamefont {J.~L.}\ \bibnamefont {Bohn}},\ }\href
  {\doibase 10.1103/PhysRevA.80.023614} {\bibfield  {journal} {\bibinfo
  {journal} {Phys. Rev. A}\ }\textbf {\bibinfo {volume} {80}},\ \bibinfo
  {pages} {023614} (\bibinfo {year} {2009})}\BibitemShut {NoStop}%
\bibitem [{\citenamefont {Lahaye}\ \emph {et~al.}(2008)\citenamefont {Lahaye},
  \citenamefont {Metz}, \citenamefont {Fr\"{o}hlich}, \citenamefont {Koch},
  \citenamefont {Meister}, \citenamefont {Griesmaier}, \citenamefont {Pfau},
  \citenamefont {Saito}, \citenamefont {Kawaguchi},\ and\ \citenamefont
  {Ueda}}]{Lahaye2009a}%
  \BibitemOpen
  \bibfield  {author} {\bibinfo {author} {\bibfnamefont {T.}~\bibnamefont
  {Lahaye}}, \bibinfo {author} {\bibfnamefont {J.}~\bibnamefont {Metz}},
  \bibinfo {author} {\bibfnamefont {B.}~\bibnamefont {Fr\"{o}hlich}}, \bibinfo
  {author} {\bibfnamefont {T.}~\bibnamefont {Koch}}, \bibinfo {author}
  {\bibfnamefont {M.}~\bibnamefont {Meister}}, \bibinfo {author} {\bibfnamefont
  {A.}~\bibnamefont {Griesmaier}}, \bibinfo {author} {\bibfnamefont
  {T.}~\bibnamefont {Pfau}}, \bibinfo {author} {\bibfnamefont {H.}~\bibnamefont
  {Saito}}, \bibinfo {author} {\bibfnamefont {Y.}~\bibnamefont {Kawaguchi}}, \
  and\ \bibinfo {author} {\bibfnamefont {M.}~\bibnamefont {Ueda}},\ }\href
  {\doibase 10.1103/PhysRevLett.101.080401} {\bibfield  {journal} {\bibinfo
  {journal} {Phys. Rev. Lett.}\ }\textbf {\bibinfo {volume} {101}},\ \bibinfo
  {eid} {080401} (\bibinfo {year} {2008})}\BibitemShut {NoStop}%
\bibitem [{\citenamefont {Kadau}\ \emph {et~al.}(2016)\citenamefont {Kadau},
  \citenamefont {Schmitt}, \citenamefont {Wenzel}, \citenamefont {Wink},
  \citenamefont {Maier}, \citenamefont {Ferrier-Barbut},\ and\ \citenamefont
  {Pfau}}]{Kadau2016a}%
  \BibitemOpen
  \bibfield  {author} {\bibinfo {author} {\bibfnamefont {H.}~\bibnamefont
  {Kadau}}, \bibinfo {author} {\bibfnamefont {M.}~\bibnamefont {Schmitt}},
  \bibinfo {author} {\bibfnamefont {M.}~\bibnamefont {Wenzel}}, \bibinfo
  {author} {\bibfnamefont {C.}~\bibnamefont {Wink}}, \bibinfo {author}
  {\bibfnamefont {T.}~\bibnamefont {Maier}}, \bibinfo {author} {\bibfnamefont
  {I.}~\bibnamefont {Ferrier-Barbut}}, \ and\ \bibinfo {author} {\bibfnamefont
  {T.}~\bibnamefont {Pfau}},\ }\href {http://dx.doi.org/10.1038/nature16485}
  {\bibfield  {journal} {\bibinfo  {journal} {Nature}\ }\textbf {\bibinfo
  {volume} {530}},\ \bibinfo {pages} {194} (\bibinfo {year}
  {2016})}\BibitemShut {NoStop}%
\bibitem [{\citenamefont {Ferrier-Barbut}\ \emph {et~al.}(2016)\citenamefont
  {Ferrier-Barbut}, \citenamefont {Kadau}, \citenamefont {Schmitt},
  \citenamefont {Wenzel},\ and\ \citenamefont {Pfau}}]{Ferrier-Barbut2016a}%
  \BibitemOpen
  \bibfield  {author} {\bibinfo {author} {\bibfnamefont {I.}~\bibnamefont
  {Ferrier-Barbut}}, \bibinfo {author} {\bibfnamefont {H.}~\bibnamefont
  {Kadau}}, \bibinfo {author} {\bibfnamefont {M.}~\bibnamefont {Schmitt}},
  \bibinfo {author} {\bibfnamefont {M.}~\bibnamefont {Wenzel}}, \ and\ \bibinfo
  {author} {\bibfnamefont {T.}~\bibnamefont {Pfau}},\ }\href {\doibase %
  10.1103/PhysRevLett.116.215301} {\bibfield  {journal} {\bibinfo  {journal}
  {Phys. Rev. Lett.}\ }\textbf {\bibinfo {volume} {116}},\ \bibinfo {pages}
  {215301} (\bibinfo {year} {2016})}\BibitemShut {NoStop}%
\bibitem [{\citenamefont {{Bisset}}\ \emph {et~al.}()\citenamefont {{Bisset}},
  \citenamefont {{Wilson}}, \citenamefont {{Baillie}},\ and\ \citenamefont
  {{Blakie}}}]{Bisset2016a}%
  \BibitemOpen
  \bibfield  {author} {\bibinfo {author} {\bibfnamefont {R.~N.}\ \bibnamefont
  {{Bisset}}}, \bibinfo {author} {\bibfnamefont {R.~M.}\ \bibnamefont
  {{Wilson}}}, \bibinfo {author} {\bibfnamefont {D.}~\bibnamefont {{Baillie}}},
  \ and\ \bibinfo {author} {\bibfnamefont {P.~B.}\ \bibnamefont {{Blakie}}},\
  }\href@noop {} {\ }\Eprint {http://arxiv.org/abs/1605.04964}
  {arXiv:1605.04964} \BibitemShut {NoStop}%
\bibitem [{\citenamefont {W\"achtler}\ and\ \citenamefont
  {Santos}(2016)}]{Wachtler2016a}%
  \BibitemOpen
  \bibfield  {author} {\bibinfo {author} {\bibfnamefont {F.}~\bibnamefont
  {W\"achtler}}\ and\ \bibinfo {author} {\bibfnamefont {L.}~\bibnamefont
  {Santos}},\ }\href {\doibase 10.1103/PhysRevA.93.061603} {\bibfield
  {journal} {\bibinfo  {journal} {Phys. Rev. A}\ }\textbf {\bibinfo {volume}
  {93}},\ \bibinfo {pages} {061603} (\bibinfo {year} {2016})}\BibitemShut
  {NoStop}%
\bibitem [{\citenamefont {Saito}(2016)}]{Saito2016a}%
  \BibitemOpen
  \bibfield  {author} {\bibinfo {author} {\bibfnamefont {H.}~\bibnamefont
  {Saito}},\ }\href {\doibase 10.7566/JPSJ.85.053001} {\bibfield  {journal}
  {\bibinfo  {journal} {J. Phys. Soc. Jpn}\ }\textbf {\bibinfo {volume} {85}},\
  \bibinfo {pages} {053001} (\bibinfo {year} {2016})}\BibitemShut {NoStop}%
\bibitem [{\citenamefont {Ronen}\ \emph {et~al.}(2006)\citenamefont {Ronen},
  \citenamefont {Bortolotti},\ and\ \citenamefont {Bohn}}]{Ronen2006a}%
  \BibitemOpen
  \bibfield  {author} {\bibinfo {author} {\bibfnamefont {S.}~\bibnamefont
  {Ronen}}, \bibinfo {author} {\bibfnamefont {D.~C.~E.}\ \bibnamefont
  {Bortolotti}}, \ and\ \bibinfo {author} {\bibfnamefont {J.~L.}\ \bibnamefont
  {Bohn}},\ }\href {\doibase 10.1103/PhysRevA.74.013623} {\bibfield  {journal}
  {\bibinfo  {journal} {Phys. Rev. A}\ }\textbf {\bibinfo {volume} {74}},\
  \bibinfo {pages} {013623} (\bibinfo {year} {2006})}\BibitemShut {NoStop}%
\bibitem [{\citenamefont {Lima}\ and\ \citenamefont
  {Pelster}(2011)}]{Lima2011a}%
  \BibitemOpen
  \bibfield  {author} {\bibinfo {author} {\bibfnamefont {A.~R.~P.}\
  \bibnamefont {Lima}}\ and\ \bibinfo {author} {\bibfnamefont {A.}~\bibnamefont
  {Pelster}},\ }\href {\doibase 10.1103/PhysRevA.84.041604} {\bibfield
  {journal} {\bibinfo  {journal} {Phys. Rev. A}\ }\textbf {\bibinfo {volume}
  {84}},\ \bibinfo {pages} {041604} (\bibinfo {year} {2011})}\BibitemShut
  {NoStop}%
\bibitem [{\citenamefont {Bisset}\ and\ \citenamefont
  {Blakie}(2015)}]{Bisset2015a}%
  \BibitemOpen
  \bibfield  {author} {\bibinfo {author} {\bibfnamefont {R.~N.}\ \bibnamefont
  {Bisset}}\ and\ \bibinfo {author} {\bibfnamefont {P.~B.}\ \bibnamefont
  {Blakie}},\ }\href {\doibase 10.1103/PhysRevA.92.061603} {\bibfield
  {journal} {\bibinfo  {journal} {Phys. Rev. A}\ }\textbf {\bibinfo {volume}
  {92}},\ \bibinfo {pages} {061603} (\bibinfo {year} {2015})}\BibitemShut
  {NoStop}%
\bibitem [{\citenamefont {Blakie}(2016)}]{Blakie2016a}%
  \BibitemOpen
  \bibfield  {author} {\bibinfo {author} {\bibfnamefont {P.~B.}\ \bibnamefont
  {Blakie}},\ }\href {\doibase 10.1103/PhysRevA.93.033644} {\bibfield
  {journal} {\bibinfo  {journal} {Phys. Rev. A}\ }\textbf {\bibinfo {volume}
  {93}},\ \bibinfo {pages} {033644} (\bibinfo {year} {2016})}\BibitemShut
  {NoStop}%
\bibitem [{\citenamefont {W\"achtler}\ and\ \citenamefont
  {Santos}()}]{Wachtler2016c}%
  \BibitemOpen
  \bibfield  {author} {\bibinfo {author} {\bibfnamefont {F.}~\bibnamefont
  {W\"achtler}}\ and\ \bibinfo {author} {\bibfnamefont {L.}~\bibnamefont
  {Santos}},\ }\href@noop {} {\ }\Eprint {http://arxiv.org/abs/1605.08676}
  {arXiv:1605.08676} \BibitemShut {NoStop}%
\end{thebibliography}

%

\end{document}